\documentstyle[epsf,axodraw]{elsart}
\journal{Physics Letters B}

\newcommand{\ee}{e^+e^-}
\newcommand{\ww}{W^+W^-}

\newcommand{\s}{\sigma }
\newcommand{\sba}{\bar\sigma }
\newcommand{\la}{\lambda }
\newcommand{\lb}{\bar\lambda }
\newcommand{\cM}{{\cal M}}

\newcommand{\beq}{\begin{equation}}
\newcommand{\eeq}{\end{equation}}
\newcommand{\nn}{\nonumber}
\newcommand{\bea}{\begin{eqnarray}}
\newcommand{\eea}{\end{eqnarray}}

\newcommand{\rfn}[1]{(\ref{#1})}
\newcommand{\Eq}[1]{Eq.(\ref{#1})}

\newcommand{\ea}{{\it et al.}}

\newcommand{\eg}{{\it e.g.}}

\newcommand{\np}[1]{Nucl. Phys. {\bf #1}}
\newcommand{\pl}[1]{Phys. Lett. {\bf #1}}
\newcommand{\pr}[1]{Phys. Rev. {\bf #1}}
\newcommand{\prl}[1]{Phys. Rev. Lett. {\bf #1}}
\newcommand{\zp}[1]{Z. Phys. {\bf #1}}
\newcommand{\epj}[1]{Eur. Phys. J. {\bf #1}}
\newcommand{\prep}[1]{Phys. Rep. {\bf #1}}

\def\lsim{\mathrel{\vcenter{\hbox{$<$}\nointerlineskip\hbox{$\sim$}}}}
\def\gsim{\mathrel{\vcenter{\hbox{$>$}\nointerlineskip\hbox{$\sim$}}}}

\begin{document}

\begin{frontmatter}

\hfill UCRHEP-T269

\title{\bf TESTING LEFT-RIGHT SYMMETRIC GAUGE THEORIES AT $e^+e^-$
           COLLIDERS}
\author{M. Raidal}  
\address{DESY, Deutsches Elektronen-Synchroton, D-22603 Hamburg, Germany,\\
Department of Physics, University of California, Riverside,
CA 92521, U.S.A., and\\
National Institute of Chemical Physics and Biophysics,
10143 Tallinn, Estonia}

\begin{abstract}
If the Standard Model is embedded in a left--right symmetric gauge theory 
at the TeV scale, the pair production of  light $W$--bosons
in $\ee$ collisions, $\ee\to\ww$, will 
be affected by mixings in the gauge and neutrino sectors, and 
by the $t$--channel exchange of a heavy right--handed neutrino. 
The modification of the cross section by 
these new effects is studied for high--energy $e^+e^-$ colliders.
\end{abstract}
\end{frontmatter}


{\bf 1.} {\it \underline{Introduction}.} 
The embedding of the Standard Model (SM) in a left--right (LR)
symmetric theory \cite{lr}
at scales of order 1 TeV is a hypothetical but interesting option.
Models based on the gauge group $SU(2)_L\times SU(2)_R\times U(1)_{B-L}$
share the success of the SM and, in addition, are able to explain
the parity violation in the weak interaction in a dynamical way.
They necessarily incorporate right--handed neutrino states which 
can be connected with the non-zero neutrino masses via the see-saw 
mechanism \cite{seesaw}. Moreover, spontaneous symmetry breaking may 
be the origin of CP--violation in LR models \cite{cp}.
Even the high left--right breaking scales are generally 
preferred in theoretical analyses\footnote{However, in supersymmetric
LR models \cite{susylr} there is an upper bound on the right-handed 
scale set by the scale of supersymmetry.}, 
a scale of order one to several
TeV is still compatible with all direct experimental observations
motivating collider searches.

\begin{figure}[t]
\begin{center}
\begin{picture}(230,100)(10,0)
\Text(5,85)[]{$e^-$}
\ArrowLine(0,75)(30,50)
\ArrowLine(30,50)(0,25)
\Text(5,20)[]{$e^+$}
\Photon(30,50)(70,50){3}{5}
\Photon(70,50)(100,75){3}{5}
\Photon(70,50)(100,25){3}{5}
\Text(50,60)[]{$\gamma,Z,Z'$}
\Text(95,82)[]{$W^-$}
\Text(95,18)[]{$W^+$}
\Text(145,85)[]{$e^-$}
\ArrowLine(140,75)(185,75)
\Text(145,35)[]{$e^+$}
\ArrowLine(185,25)(140,25)
\ArrowLine(185,75)(185,25)
\Text(200,50)[]{$\nu,N$}
\Photon(185,75)(230,75){3}{5}
\Text(225,85)[]{$W^-$}
\Text(225,35)[]{$W^+$}
\Photon(185,25)(230,25){3}{5}
\end{picture}
\end{center}
\caption{\it 
Diagrams contributing to the process $ee\to\ww$ in the LR model.
\vspace*{0.5cm} }
\label{fig:eeww}
\end{figure}
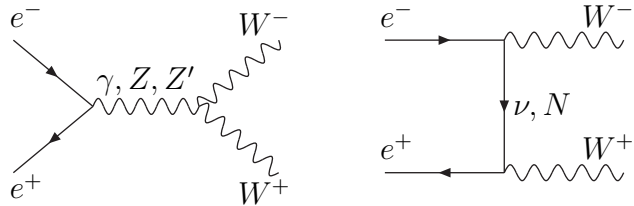

Consequences of  LR extensions of the SM have been studied
in many facets of the theory. In this letter we shall focus on the 
impact of such extensions on the pair production of the light $W$--bosons 
at $\ee$ colliders:
\bea
\ee\to\ww ,
\label{eeww}
\eea 
a precursor to the production of the new heavy charged gauge bosons
which may require collider energies in the multi-TeV range.
This process is affected in LR  symmetric theories by three mechanisms.
First, the interpretation of the observed 
mass eigenstates $W^\pm$ as mixed states of 
$W^\pm_L$ with a small admixture of $W^\pm_R$, as well as 
the interpretation of the light electron neutrino state $\nu$ 
as a mixture of the left--handed neutrino state $\nu_L$ with the heavy 
right--handed neutrino state $\nu_R.$
Second, the $t$--channel exchange of the new heavy, predominantly 
right--handed neutrino $N$, cf. Fig.\ref{fig:eeww}. Third,
the mixing in the neutral gauge boson sector 
and the $s$--channel exchange of the heavy 
neutral gauge boson $Z'.$ 
These three mechanisms modify the total cross section, 
the angular distributions
as well as the $W$ helicities in the final states. 
Such effects have been searched for at LEP2 \cite{5a}, 
and they will be searched for \cite{leike}  
at future $\ee$ linear colliders \cite{accomando}.
If not discovered, new limits may be derived on the LR parameters.
This note expands the earlier analysis of Ref.\cite{pav} by
taking properly inot account the mixings in the charged $W$-boson sector.
The effects are in general different from the 
effects induced by the anomalous gauge boson self--interactions.
Other LR analyses in $\ee$ collisions focussed on the exchange of 
heavy $W'^\pm$ states \cite{god} and on the Higgs phenomena \cite{hui}.

%


\vskip 1cm

{\bf 2.} {\it \underline{The Model}.}
In the minimal $SU(2)_L \times SU(2)_R \times U(1)_{B-L} $ model 
each generation of quarks and leptons carry the quantum numbers 
$Q_L\sim (1/2,0,1/3),$ $Q_R\sim (0,1/2,1/3),$ $L_L\sim (1/2,0,-1)$ and
$L_R\sim (0,1/2,-1).$ The right--handed fields are doublets 
under $SU(2)_R$ and a right--handed neutrino $N_R$ must exist. 
The minimal Higgs sector consists of a bidoublet $\phi\sim (1/2,1/2,0)$
and two triplest $\Delta_L \sim (1,0,2)$ and $ \Delta_R \sim (0,1,2).$
After the spontaneous symmetry breaking, the phenomenological
requirement $ |v_R| \gg |k_1|, |k_2|  \gg |v_L| $ 
for the vacuum expectation values $v_{L,R}$ and $k_{1,2}$ of the triplet and 
doublet Higgs fields, ensures the suppression of the 
right--handed currents and  the smallness of the 
neutrino mass.

The  $SU(2)_L \times SU(2)_R \times U(1)_{B-L} $ gauge 
symmetry group implies that
the usual left--handed gauge bosons $W_L^i$ $(i=1,$ 2, 3), 
their right--handed counterparts $W_R^i$ and the $U(1)$ gauge boson $Y$
combine to form the physical charged and neutral
gauge bosons and the photon. In general, the strength of the 
gauge interactions of these bosons is described by 
the  coupling constants $g_L,$ $g_R$ and $g',$ respectively. 
However, strict LR symmetry 
$ \Psi_L \leftrightarrow \Psi_R, 
\Delta_L \leftrightarrow \Delta_R,
\phi \leftrightarrow \phi^\dagger $
[with $\Psi$ denoting any fermion] leads to the relation $g_L=g_R\equiv g,$
which will be assumed throughout this paper.

The weak eigenstates $W_L^\pm$ and $W_R^\pm$ 
mix in the mass eigenstates $W^\pm$ and $W'^\pm$; assuming CP invariance,
the mixing matrix is defined by the angle $\chi_W$:
\bea
W^{\pm} & \, = \, & \; \; \; \cos\chi_W W^{\pm}_L \, + \,\sin\chi_W W^{\pm}_R 
\nn \\
W'^{\pm} &\,  = \, & -\sin\chi_W W^{\pm}_L\,+\,\cos\chi_W W^{\pm}_R 
\eea
The weak eigenstate $W_L$ can be identified with the pure 
SM gauge boson.
Similarly the neutrino mass eigenstates are mixtures of the weak eigenstates,
parametrized by the angle $\chi_N$:
\bea
\nu  \, &  = & \;\;\;\; \cos \chi_N\, \nu'\, +\, \sin\chi_N\, N' \nn \\
N \, & = & \, -\sin\chi_N\, \nu'\, + \, \cos\chi_N\, N'
\eea
 $\nu$ and $N$ are the light and heavy neutrino mass eigenstates,
and  $\nu'=\nu_L+\nu_L^c$ and  
$N'=\nu_R+\nu_R^c$ are the usual self-conjugate spinors; 
intergenerational mixings are irrelevant for the present analyses.

The charged--current interaction vertices for the left--chiral and 
the right--chiral currents are given by 
\bea
\langle \nu_L | W_L |e^-\rangle &=& \frac{g}{2\sqrt{2}}  
W^{\dagger\mu}_L  \, \overline{\nu_L} \, \gamma_\mu \,(1\,-\,\gamma_5) \, e 
  \nn\\
\langle N_R | W_R |e^-\rangle &=& \frac{g}{2\sqrt{2}}  
W^{\dagger\mu}_R  \, \overline{N_R} \, \gamma_\mu \,(1\,+\,\gamma_5) \, e 
\label{cc}
\eea
The charged--current interactions for the mixed mass eigenstates can 
easily be obtained from these matrix elements.

The neutral gauge bosons in LR models are mixtures of $W^3_{L,R}$ and $Y.$ 
The mixing between the massive neutral gauge bosons 
relevant for our analyses can be parametrized  as  
\bea
Z \, & = & \, \; \; \; \cos\chi_Z \, Z_1 \, + \, \sin\chi_Z \, Z_2 \nn \\
Z' \, & = & \, -\sin\chi_Z \, Z_1 \, + \, \cos\chi_Z \, Z_2
\eea
where  $Z$ and $Z'$ denote the
mass eigenstates, and $Z_1$ and $Z_2$ denote the
weak eigenstates of the massive neutral bosons.
The field $Z_1$ can be identified as the corresponding SM boson.

The tree-level neutral current Lagrangian for the 
physical $Z,Z'$ bosons is of the form
\bea
{\cal L}_{NC}  &=&
\frac{g}{2\cos\theta_W}\left[\bar f\gamma_\mu
\left( g_V^f-g_A^f\gamma_5\right) f\; Z_\mu + 
\bar f\gamma_\mu
\left( g_V^{\prime f}-g_A^{\prime f}\gamma_5\right) f\; Z'_\mu
\right]
\label{treenc}
\eea
where
\bea
g_V^f&=&\cos{\chi_Z}\;g_{V}^{0f}+\sin{\chi_Z}\;g_{V}^{\prime f} \, \nn \\
g_A^f&=&\cos{\chi_Z}\;g_{A}^{0f}+\sin{\chi_Z}\;g_{A}^{\prime f} \, 
\eea
and
\bea
\begin{array}{lll}
g_{V}^{0f} = I_{3}^f-2Q^f\sin^2\theta_{W}\,, &\;\;\;\;\;&  
g_{V}^{\prime f}= 1/\sqrt{\cos 2\theta_W }\; g_{V}^{0f}  \nn \\
g_{A}^{0f} = I_{3}^f \,, &\;\;\;\;\;&  
g_{A}^{\prime f} = -\sqrt{\cos 2\theta_W }\; g_{A}^{0f}  
\end{array}
\label{gv0}
\eea

Here we have taken into account that 
in the LRSM always $I_{3R}^f=I_{3L}^f\equiv I_3^f$ 
for the third components of the L/R isospin for a given fermion flavor $f.$ 
For simplicity, we have expressed the couplings in terms of 
$I_{3}^f$ and the electric charge $Q_f.$

Elaborate analyses of high-precision data have been presented in the 
past years on the LR gauge sector \cite{zprim,zralek}, constraining 
the $Z'$ mass to values above 
${\cal O}(1)$ TeV scale and the mixing 
among the neutral gauge bosons to values 
below ${\cal O}(10^{-4})$. In the present context
these effects in the neutral currents are 
expected to be subleading compared with the possible effects in the 
charged current interactions.

The lower bound on the $W'$ mass derived from the $K_L$-$K_S$
mass difference is quite stringent \cite{beall}, $M_{W'}\gsim 1.6$ TeV
(being, however,  subject to uncertainties from low energy QCD 
in the kaon system); the bound on the mixing angle $\chi_W$ 
is as low as $\chi_W\lsim 0.013 $ \cite{lan}.
The direct searches for $W'$ at the Tevatron yield 
bounds $M_{W'}\gsim 720$ GeV
assuming a light keV--range $N,$ and $M_{W'}\gsim 650$ GeV assuming
$M_N<M_{W'}/2$ \cite{tev}. These bounds are weakened considerably 
for  more general LR models \cite{rizzo}.

The least tested components of the LR model are the masses and mixings 
of  neutrinos. Analyses of the precision data that constrain 
fermion mixings \cite{esteban} have given a
90\% CL bound $|\chi_N|\lsim 0.081$ for the electron neutrinos.
As no new particles have been discovered at LEP2 it is 
plausible to assume that the mass of the heavy R--type neutrino 
may exceed about 100 GeV.

The mixing angle $\chi_W$ is, even in the simplest models of the Higgs
representations, independent of the neutral current parameters.
Also the prediction on the mixing angle $\chi_N$
depends strongly on the scenario in which the neutrino mixings and masses are 
generated \cite{BR}, allowing  mixings still as large as  
$\chi_N\sim 0.1$.


\vskip 1cm

{\bf 3.} {\it \underline{Pair Production of W-Bosons}.}
In the LR symmetric theory the process $\ee\to\ww$ is 
built up by $s$--channel exchanges of $\gamma,$ $Z$ and $Z'$
bosons and $t$--channel neutrino $\nu$ and $N$ exchanges, cf. 
Fig.\ref{fig:eeww}. As a result, the cross section and the distributions in 
the process $\ee\to\ww$ are modified by $W$ and $\nu$ mixings 
and $N$ exchange,
and by the mixings in the neutral boson sector and the $Z'$ exchange.

The helicity amplitudes of the process \rfn{eeww} in the LR model
can be written as
\bea
\cM (\s,\sba;\la,\lb)=\cM_\gamma +\cM_{Z}+\cM_{Z'} +\cM_\nu +  \cM_N 
\eea
where $\s/2$ and $\sba/2$ denote the helicities of the incoming electron and 
positron, respectively; 
$\la$ and $\lb$ are the helicities of the  $W^-$ and $W^+$ bosons,
 respectively.
In the following analysis we closely follow the notation of Ref.\cite{cern}
by writing the helicity amplitudes in terms of reduced
amplitudes $\tilde \cM$ which are defined by the relation
\bea
\cM_X (\s,\sba;\la,\lb|\theta)=
\sqrt{2}\;\s\; e^2\;\tilde\cM_{X}(\s,\sba;\la,\lb|\theta)
\;d^{J_0}_{\Delta\s,\Delta\la}(\theta)\;
\label{redamp}
\eea 
where $X=\gamma,\,Z,\,Z',\,\nu,\,N$ and 
\bea
\Delta\s = \frac{1}{2}(\s-\sba)\;,\;\;\;\;\; 
 \Delta\la = \,(\la-\lb)\;,\;\;\;\;\;\; J_0=max(|\Delta\s|,|\Delta\la|)\;  
\eea
The angle $\theta$ is the $W^-$ production angle with respect to the 
electron momentum. 
The function $d^{J_0}_{\Delta\s,\Delta\la}(\theta)$ denotes 
the  angular--momentum wave--function associated with the minimum
angular momentum $J_0$ in the production process.
The relevant $d$ functions are collected in Table 1 which  extends
the corresponding table of Ref.\cite{cern}.

{\bf a)} For the $W^\pm$ helicity combinations $|\la - \lb| = 2$, {\it i.e.} 
$(\la,\lb)=(+,-)$ and $(-,+)$,  only $t$-channel neutrino 
exchanges contribute.
The reduced helicity amplitudes are given
by
\bea
\tilde\cM_\nu (\s=-\sba=-;\la=-\lb=\pm)&=&
-\frac{\sqrt{2}}{\sin^2\theta_W} 
\frac{\cos^2\chi_W\cos^2\chi_N}{(1+\beta^2-2\beta\cos\theta)}
 \nn \\
\tilde\cM_N (\s=-\sba=-;\la=-\lb=\pm)&=&
-\frac{\sqrt{2}}{\sin^2\theta_W} 
\frac{\sin^2\chi_N}{(1+\beta^2-2\beta\cos\theta + \mu_N^2)}
 \nn \\
\tilde\cM_N (\s=-\sba=+;\la=-\lb=\pm) &=&
\frac{\sqrt{2}}{\sin^2\theta_W}
\frac{\sin^2\chi_W}{(1+\beta^2-2\beta\cos\theta + \mu_N^2)}
 \nn \\
\tilde\cM_N (\s=\sba=\pm;\la=-\lb=\pm)&=& 0\,
\eea
The $W$ velocity is given by $\beta=\sqrt{1-4M^2_{W}/s}$ 
and $\mu_N=2M_N/\sqrt{s}$ is the scaled mass parameter of the heavy neutrino.
The amplitude $\tilde\cM_\nu$ describes the light neutrino exchange which 
is reduced  by the coefficient $\cos^2\chi_W\cos^2\chi_N$
compared with the SM amplitude [{\it i.e.} $\sim (1-\chi_W^2-\chi_N^2)$ 
in the leading order of the mixing angles].
 The amplitudes $\tilde\cM_N$ are generated by the heavy neutrino exchange,
characteristic of the LR model. 
The amplitude $\tilde\cM_N(\s=-,\sba=+)$ 
is induced by the left--handed interaction 
in both $W N e$ vertices,  suppressed by the neutrino
mixing angle squared $\sin^2\chi_N.$
The amplitude $\tilde\cM_N (\s=+,\sba=-),$ by contrast, is induced by 
the right--handed interaction in both charged--current vertices, 
suppressed by the mixing angle squared $\sin^2\chi_W.$ 
The amplitude $\tilde\cM_N (\s=\sba)$ for equal electron/positron 
helicities is built up by a mixture of  left-- and right--handed
interactions, which vanishes for $|\Delta\la|=2.$

{\bf b)} For $|\la-\lb|=0,1$ also $s$--channel exchanges 
of vector bosons are possible.
The corresponding reduced amplitudes can be written as
\bea
\label{subamp}
\tilde\cM_\gamma (\s=-\sba=\pm;\la,\lb)
&=& -\beta A_{\la\lb}\, \nn \\
\tilde\cM_{Z} (\s=-\sba=\pm;\la,\lb)
&=& \beta A_{\la\lb}
\left[\left(1-\frac{\sin^2\chi_W}{\cos^2\theta_W}\right)\cos\chi_Z +
\frac{\sqrt{\cos 2\theta_W}}{\cos^2\theta_W}\sin^2\chi_W\sin\chi_Z 
\right] \nn\\
&& 
\left[\cos\chi_Z\left(1-\delta_{\s,-1}\frac{1}{2\sin^2\theta_W}\right)+ 
\right. \nn\\
&& \left. \frac{\sin\chi_Z}{\sqrt{\cos 2\theta_W}} \left( 
 1-\frac{1}{2}\delta_{\s,-1}-
\delta_{\s,+1}\frac{\cos^2\theta_W}{2\sin^2\theta_W}
\right)\right]   \frac{s}{s-M^2_{Z}} \, \nn \\ 
\tilde\cM_{Z'} (\s=-\sba=\pm;\la,\lb)
&=& \beta A_{\la\lb}
\left[\left(\frac{\sin^2\chi_W}{\cos^2\theta_W}-1 \right)\sin\chi_Z +
\frac{\sqrt{\cos 2\theta_W}}{\cos^2\theta_W}\sin^2\chi_W\cos\chi_Z  
\right] \nn\\
&& \frac{\cos\chi_Z}{\sqrt{\cos 2\theta_W}} \left( 
 1-\frac{1}{2}\delta_{\s,-1}-
\delta_{\s,+1}\frac{\cos^2\theta_W}{2\sin^2\theta_W}
\right)  \, \frac{s}{s-M^2_{Z'}} \nn \\
\tilde\cM_\nu (\s=-\sba=-;\la,\lb)
&=&
\frac{\cos^2\chi_W\cos^2\chi_N}{2\beta\sin^2\theta_W} 
\left[ B_{\la\lb}- \frac{1}{1+\beta^2-2\beta\cos\theta} 
C^\nu_{\la\lb}\right] \, \nn \\
\tilde\cM_N  (\s=-\sba=-;\la,\lb)
&=&
\frac{\sin^2\chi_N}{2\beta\sin^2\theta_W} 
\left[ B_{\la\lb}- \frac{1}{1+\beta^2-2\beta\cos\theta+\mu_N^2} 
C^{N}_{\la\lb}\right] \, \nn \\
\tilde\cM_N   (\s=-\sba=+;\la,\lb)
&=& -\frac{\sin^2\chi_W}{2\beta\sin^2\theta_W} 
\left[ B_{\la\lb}- \frac{1}{1+\beta^2-2\beta\cos\theta+\mu_N^2} 
C^{N}_{\la\lb}\right] \,   \\
\tilde\cM_N(\s=\sba=\pm;\la,\lb)
&=& 
\frac{\sqrt{2}\sin\chi_N\sin\chi_W}{2\beta\sin^2\theta_W} 
\mu_N
\left[ D_{\la\lb}- \frac{1}{1+\beta^2-2\beta\cos\theta+\mu_N^2} 
E_{\la\lb}\right] \, \nn
\eea
The subamplitudes $A_{\la\lb}$ to $E_{\la\lb}$ are collected in Table 2.

\begin{table}
\caption{Angular momentum
$d$--functions of the process $e^+e^-\to W^+W^-$ 
as defined in \Eq{redamp}; $\eta_\sigma=sign(\Delta\s)$
and $\eta_\la=sign(\Delta\la).$
\vspace*{0.3cm} }
\begin{center}
$
\begin{array}{|c|c|c||c|}
\hline
\;|\,\Delta\s\,|\; &\; |\,\Delta\la\,|\; & \;\;J_0\;\; & 
d^{J_0}_{\Delta\s,\Delta\la} \\
\hline\hline
 1 & 2 & 2 &\; \eta_\la (1+\eta_\s\eta_\la\beta)\sin\theta/2\; \\
 1 & 1 & 1 &  (1+\eta_\s\eta_\la\beta)/2 \\
 1 & 0 & 1 & -\eta_\s \sin\theta/\sqrt{2} \\
 0 & 2 & 2 & 0 \\
 0 & 1 & 1 &  \sin\theta/\sqrt{2} \\
 0 & 0 & 0 & 1 \\
\hline
\end{array}
$
\end{center}
\end{table}

\begin{table}
\caption{Subamplitudes 
of the process $e^+e^-\to W^+W^-$ as defined in \Eq{subamp}
for the $W$ helicities $|\la-\lb |=0,1$.
The parameter $\gamma$ is the boost factor $\gamma=\sqrt{s}/2M_W$. 
\vspace*{0.3cm}}
\begin{center}
$
\begin{array}{|c||c|c|c|c|c|c|}
\hline
\lambda\bar\lambda & A_{\lambda\bar\lambda} & 
B_{\lambda\bar\lambda} & C^\nu_{\lambda\bar\lambda} &
C^N_{\lambda\bar\lambda} &
D_{\lambda\bar\lambda} & E_{\lambda\bar\lambda} \\
\hline\hline
\pm\pm & 1 & 1 & 1/\gamma^2 & 1/\gamma^2+\mu_N^2 &
\la/2  & \la[(1+\la\s\beta)^2+\mu_N^2]/2  \\
\pm 0 & 2\gamma & 2\gamma & 2(1+\la\beta)/\gamma &
2[(1-\la\s\beta)/\gamma + \mu_N^2\gamma] & 
0 & \gamma\beta (1+\la\s\beta)   \\
0 \pm & 2\gamma & 2\gamma & 2(1-\lb\beta)/\gamma &
2[(1+\lb\s\beta)/\gamma + \mu_N^2\gamma] & 
0 & -\gamma\beta (1+\lb\s\beta)   \\
00 & 1+2\gamma^2 & 2\gamma^2 & 2/\gamma^2 & 
 2(1/\gamma^2+\mu_N^2\gamma^2) & 
\s\beta\gamma^2  & \s\beta\gamma^2\mu_N^2  \\
\hline
\end{array}
$
\end{center}
\vspace*{0.5cm}
\end{table}

Evidently, the process  $e^+e^-\to W^+W^-$ 
in the LR symmetric model receives non-zero contributions from all
combinations of the electron/positron helicities while the electron/positron 
helicities are always opposite in the SM. The photonic $s$--channel
contribution is not modified in the LR model. 
The $Z'$ exchange effect is small because it is suppressed both by 
heavy propagator and small mixing effects. While the $s$--channel 
$Z$ boson contribution is changed by an overall mixing factor,   
the $t$--channel neutrino contribution receives non-trivial
new contributions from the exchange of the heavy neutrino. 
The angular distributions of the final state
$W$-bosons are therefore  modified for all $W^+W^-$ helicity combinations.
Thus, not only the total cross section but also the angular distributions
of the $W$-bosons and their decay products are modified in the LR models.


\vskip 1cm

{\bf 4.} {\it \underline{Phenomenological Analyses}.}
To exemplify the results, we analyze the  LR contribution to the
total cross section of the process $\ee\to\ww.$ 
It is important to realize that standard analyses of anomalous 
static $W^\pm$ parameters cannot be applied due to the
heavy neutrino exchange in the $t$--channel which affects
angular momentum states $|J_z|\neq 1.$ The total cross section is 
given by the differential cross section integrated over the 
production angle $\theta$ in the following form:
\bea
\s[\ee\to\ww]=\frac{\pi\alpha^2}{4s}\,\beta\,\int d\cos\theta
\sum_{\s,\sba,\la,\lb} |\tilde\cM(\s,\sba;\la,\lb|\theta)\;
d^{J_0}_{\Delta\s,\Delta\la}|^2
\eea
where $\alpha\approx 1/128$ is the electromagnetic coupling evaluated at 
the high--energy scale $s.$

By adopting the estimate $\chi_Z \lsim 10^{-4}$ for $M_{Z'}=1$ TeV,
as required by precision data \cite{zprim},
the effect of $Z-Z'$ mixing on the cross section is very small,
{\it i.e.} at the per-mille level, and it can in general be neglected.

\begin{figure}[t]
\centerline{
\epsfxsize = 0.5\textwidth \epsffile{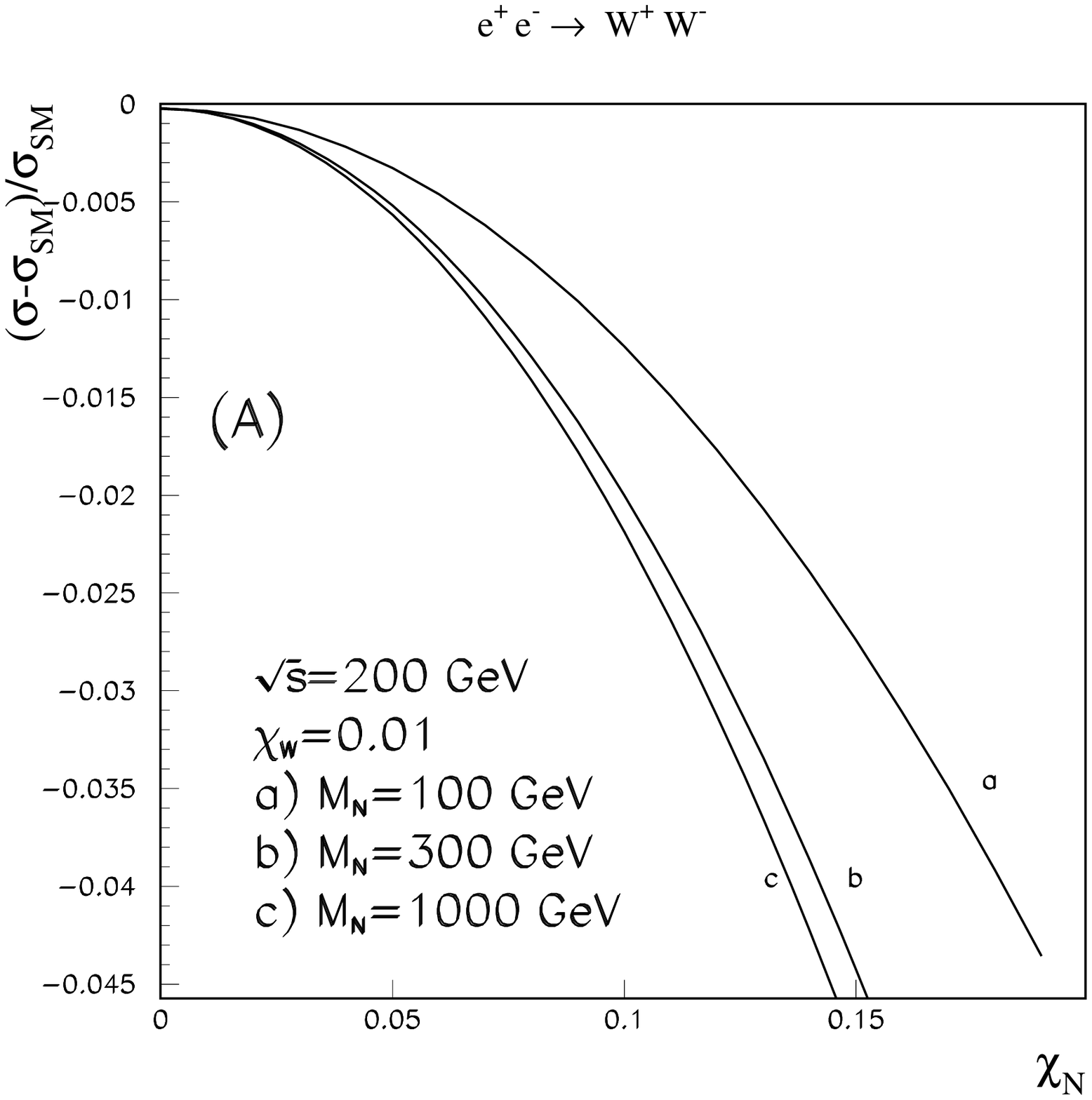} 
\hfill
\epsfxsize = 0.5\textwidth \epsffile{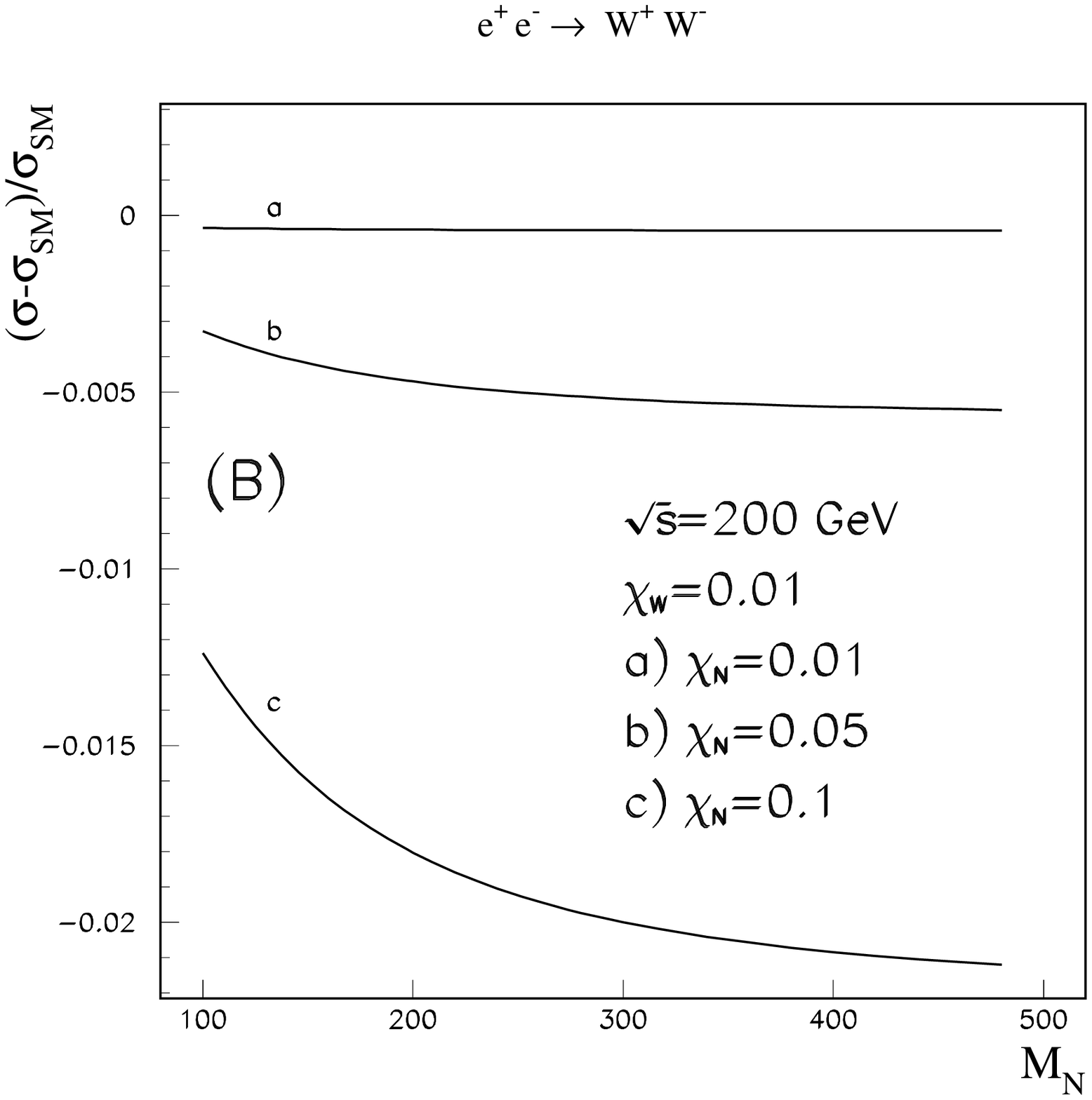}
}
\caption{\it The deviation of the unpolarized cross section of the process
$\ee\to \ww$ at LEP2 from the SM prediction as a functions of the 
mixing angle $\chi_N$ [figure (A)] and the neutrino mass $M_N$ [figure (B)]. 
The chosen numerical values are shown in the figure.
\vspace*{0.5cm}}
\label{fig:creeww}
\end{figure}

\begin{figure}[t]
\centerline{
\epsfxsize = 0.5\textwidth \epsffile{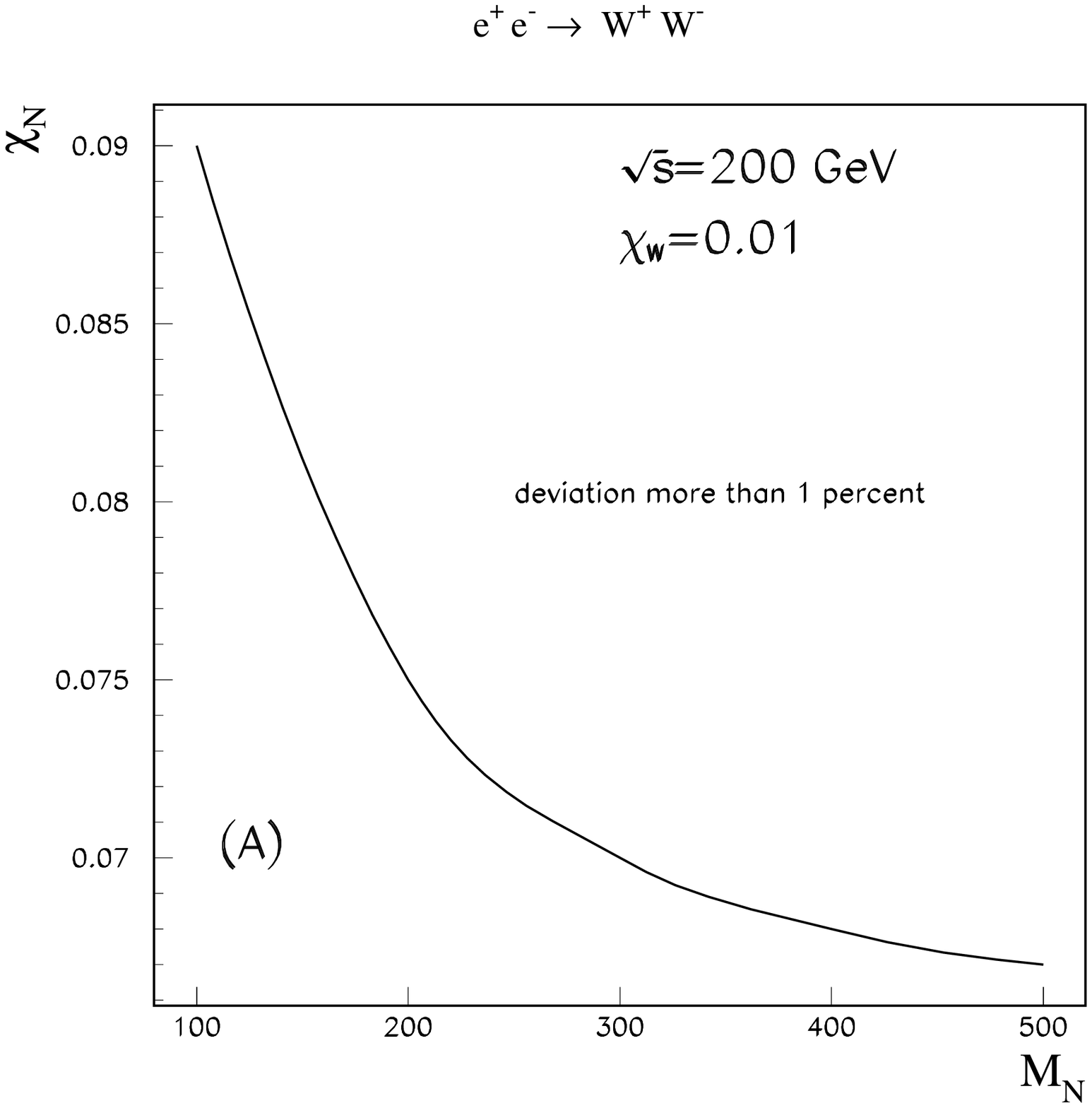} 
\hfill
\epsfxsize = 0.5\textwidth \epsffile{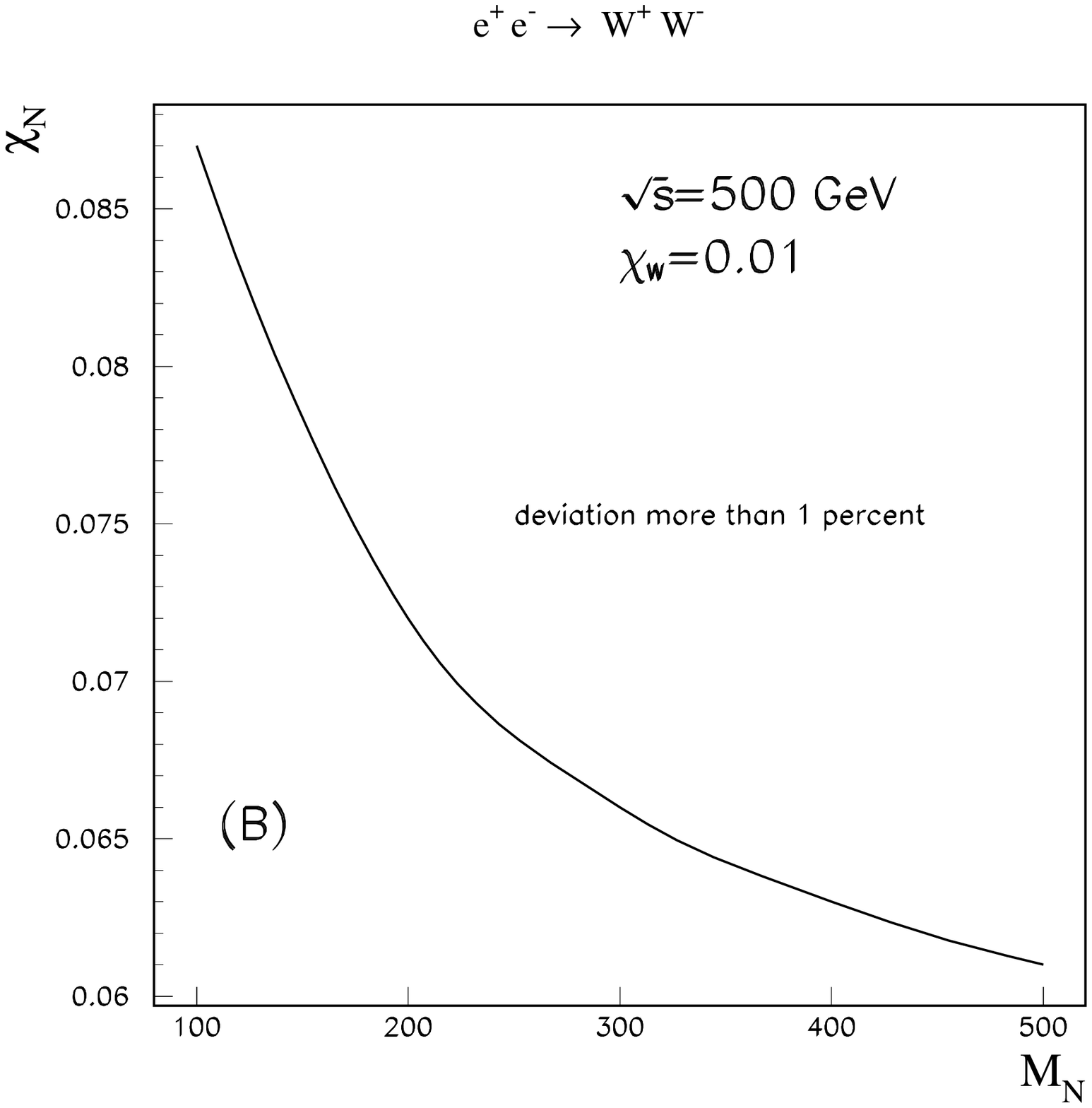} 
}
\caption{\it 
The sensitivity curves for (A) LEP2 and (B) a linear collider 
of energy $\sqrt{s}=500$ GeV.
In the region above the curve the total cross section
of the process $\ee\to\ww$ deviates from the SM prediction
by more than 1\%. Note the non--trivial behavior of the contours
which is a consequence of the counteracting mixing 
and neutrino--exchange effects. 
\vspace*{0.5cm}}
\label{fig:eewwex}
\end{figure}

The deviation from the SM cross section scaled with respect
to the improved Born approximation, is exemplified
in Figs.\ref{fig:creeww} for LEP2;
in (A) as a function of the mixing angle $\chi_N$ for fixed $M_N$, and 
in (B) as a function of the heavy neutrino mass $M_N$ 
for fixed $\chi_N.$  For the sake of simplicity 
$\sin\chi_W$ is taken $10^{-2}$ in these numerical examples.
The collision energy is set to $\sqrt{s}=200$ GeV and the values
of the parameters are indicated in the figure.
Evidently, the $WW$ production cross section is reduced
for this set of the parameters compared with the SM.
The 1\% level of the deviation from the SM is reached for 
the  mixing angles $\sin\chi_N\sim 0.07.$ 
While the quadratic dependence of $(\sigma-\sigma_{SM})/\sigma_{SM}$
on $\sin\chi$ is strong, 
the dependence on $M_N$ is rather weak.
The  effect of the non-zero mixing angle
and the heavy neutrino exchange cancel each other partly.
While non-zero mixing angles decrease the cross section, the neutrino $N$
exchange shifts the cross section upward.
This correlation is also evident from Fig.\ref{fig:eewwex} (A). 
The area  on the 
$[M_N,\sin\chi_N]$ plane in which the 
$\ww$ production cross section of the LR model 
deviates by more than 1\% from the cross section of the Standard Model
lies above the solid curve.
As a result, for relatively light right--handed neutrino masses
larger mixing angles are allowed than for heavy masses.

\begin{figure}[t]
\centerline{
\epsfxsize = 0.5\textwidth \epsffile{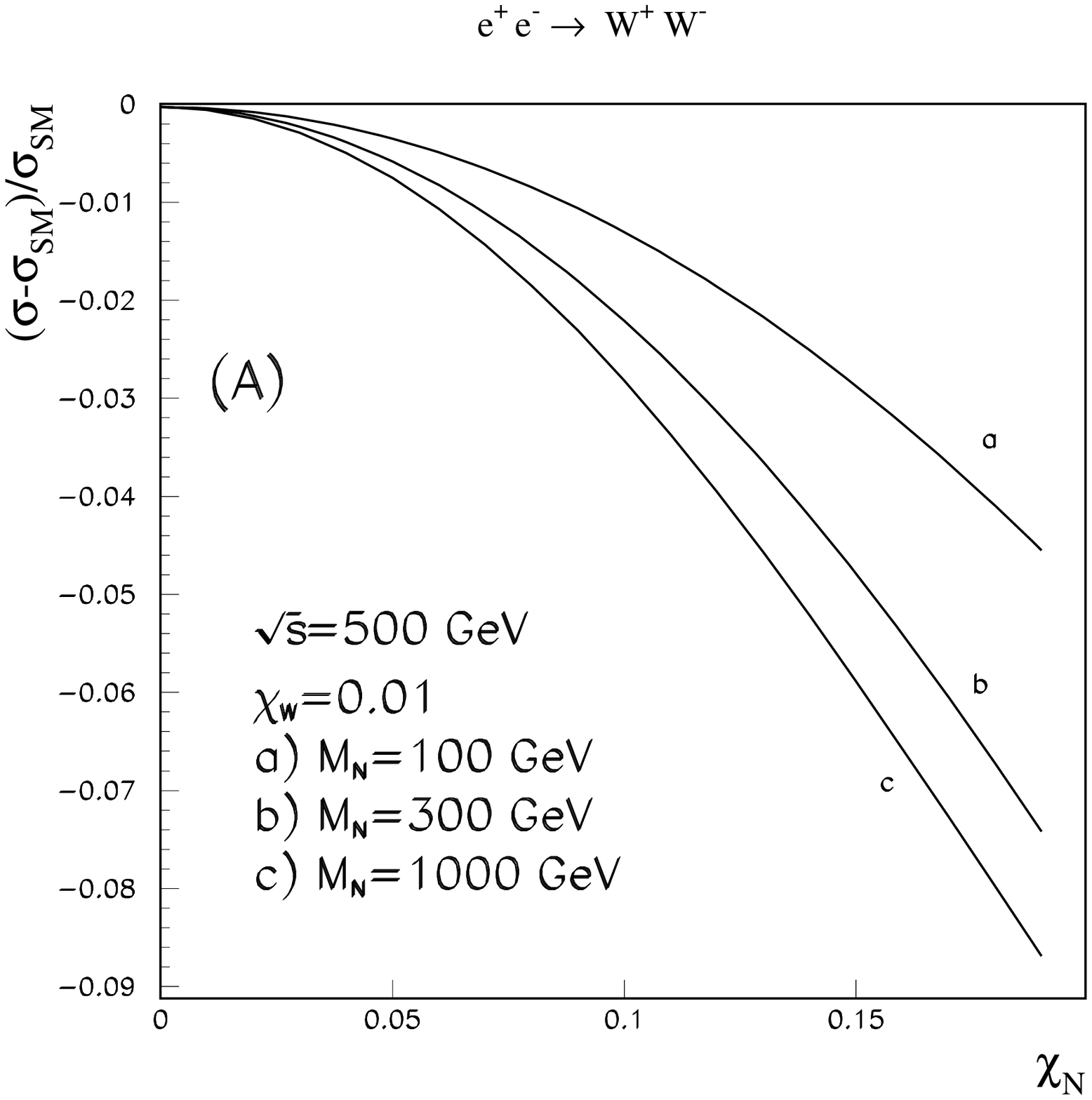} 
\hfill
\epsfxsize = 0.5\textwidth \epsffile{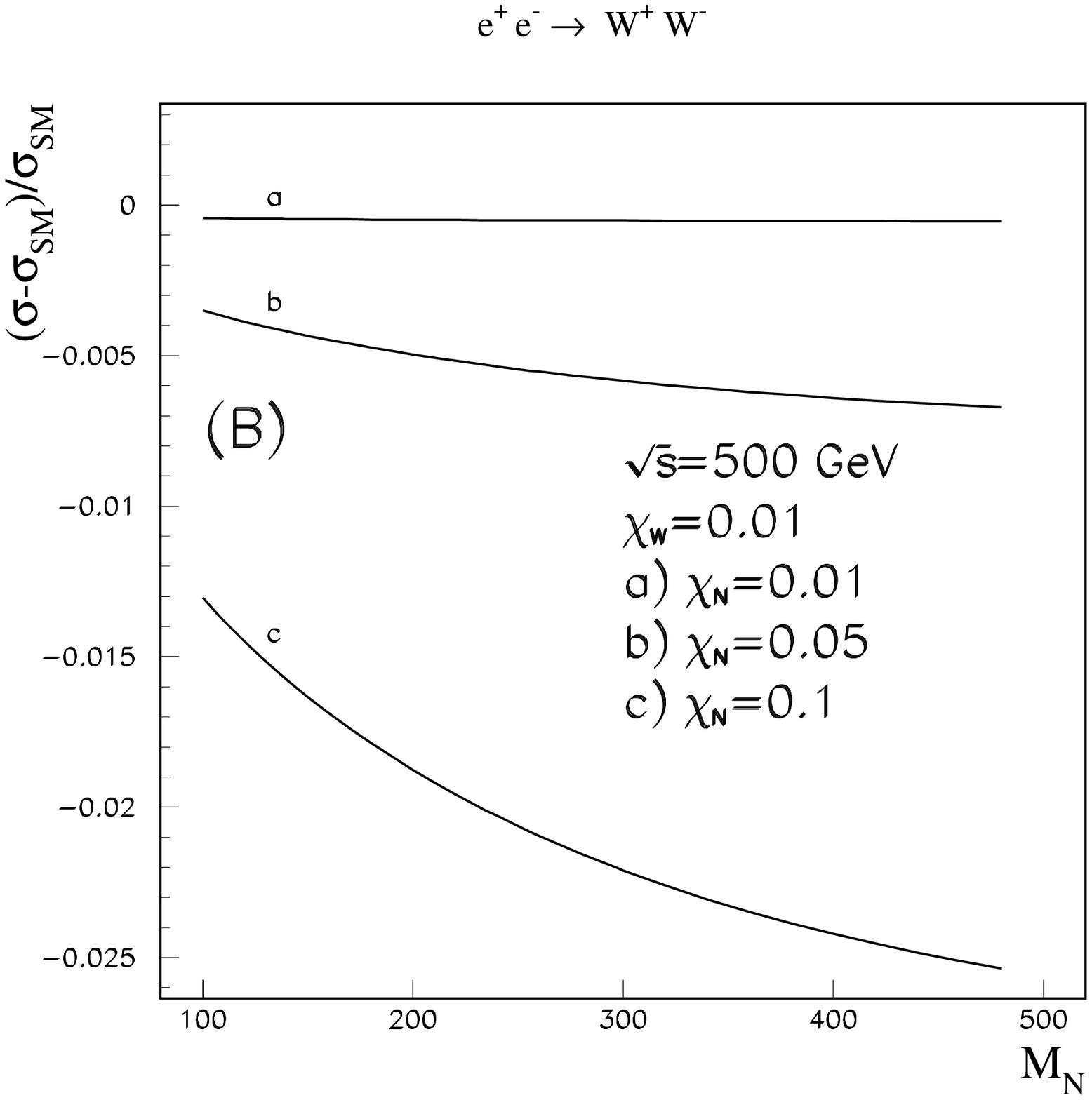}
}
\caption{\it The same analysis
as in Fig. \ref{fig:creeww} for the collision 
energy $\sqrt{s}=500$ GeV.
\vspace*{0.5cm}}
\label{fig:creeww2}
\end{figure}

We have repeated 
the analyses for a linear collider energy $\sqrt{s}=500$ GeV
in Fig.\ref{fig:creeww2}. 
While for a relatively light right--handed neutrino $N$ the increase
in the experimental sensitivity is not remarkable, for a
heavy neutrino $N$, by contrast,  
a high-energy linear collider will test the neutrino mixing angles
with significantly higher precision than LEP2. The same conclusion follows
also from Fig.\ref{fig:eewwex} (B). The reason for this behavior is 
the cancellation of the neutrino mixing effect by the heavy neutrino
exchange. If the latter is minimized at high neutrino mass, 
a linear collider will probe the mixing angle $\chi_N$ with high accuracy.

Linear colliders may also have $e^-\gamma$ collision mode in addition to 
the usual $e^+e^-$ one. Because in  $e^-\gamma$ collisions the initial 
photon can test directly the trilinear gauge boson coupling \cite{eg}, 
this collision mode will be ideal to discriminate the new physics 
discussed in this work from the anomalous gauge boson couplings. 

The analysis can be improved by exploiting the 
angular distributions of the $W$-bosons and their decay products;
significant improvements can be expected in the
region of small neutrino mass $M_N.$ Such an analysis which depends more 
strongly on the experimental conditions, 
is beyond of the scope of the present letter.


{\bf 5.} {\it \underline{Conclusions}.} 
We have shown that given the present experimental
bounds on the $W -W',$  $\nu - N$ mixings and on the $N$ mass,
observable deviations from the SM cross section and distributions
of the process $\ee\to\ww$ at $e^+e^-$ colliders
are possible in LR symmetric models. 
The dominant new effects are associated with the $t$-channel
exchange of neutrinos which differ from the anomalous
self--interaction effects of the electroweak gauge bosons. 
Independent analyses of the 
angular distributions of the $W$ boson helicity components
are needed to explore these new  effects.
Since no major deviations from the SM have been found at LEP2, 
new bounds on mixing angles and on the heavy neutrino
mass can be derived from the data. High precision experiments
at future $\ee$ linear colliders in the TeV range will extend 
the sensitivity to heavy neutrinos into the multi--TeV
region according to the scaling law $M_N\sim{s^{1/2}}\cdot{{\cal L}^{1/4}}$
for the energy squared $s$ and the integrated luminosity ${\cal L}$;
for $\sqrt{s}\sim 1$ TeV and ${\cal L}\sim 1$ ab$^{-1}$ this 
corresponds to an increase by more than an order of 
magnitude compared to LEP2.


\begin{ack}
I would like to thank P. Zerwas with whom most of this work has been done.
I am grateful to DESY Theory Group for hospitality.
His work is partly supported by the U.S. Department of Energy
under Grant No. DE-FG03-94ER40837. 
\end{ack}

\end{document}